\documentclass[superscriptaddress, nobibnotes, reprint,floatfix,%linenumbers,
]{revtex4-2}
\usepackage{upgreek}
\usepackage{amsmath,braket}
\usepackage{graphicx}
\usepackage{mathtools}
\usepackage{color}
\usepackage{bm}
\usepackage{xcolor}
\usepackage{setspace}
\usepackage{lineno}
\usepackage{ulem}
\newcommand{\RNum}[1]{\uppercase\expandafter{\romannumeral #1\relax}}
\usepackage{comment}
\usepackage{soul}
\renewcommand{\textcolor}[2]{#2}

\newcommand{\xj}{\textcolor{cyan}}
\newcommand{\zy}{\textcolor{red}}

\begin{document}

\title{Bell Inequality Violation with Vacuum-One-Photon Number Superposition States}

\author{Zi-Qi Zeng}\thanks{These authors contributed equally to this work.}
\affiliation{Beijing Academy of Quantum Information Sciences, Beijing 100193, China}
\affiliation{Beijing National Laboratory for Condensed Matter Physics, Institute of Physics, Chinese Academy of Sciences, Beijing 100190, China}
\affiliation{University of Chinese Academy of Sciences, Beijing 100049, China}

\author {Jian Wang}\thanks{These authors contributed equally to this work.}
\affiliation{Beijing Academy of Quantum Information Sciences, Beijing 100193, China}
\author{Xiu-Bin Liu}
\affiliation{Beijing Academy of Quantum Information Sciences, Beijing 100193, China}
\affiliation{School of Physics and Beijing Key Laboratory of Opto-electronic Functional Materials and Micro-nano Devices, Key Laboratory of Quantum State Construction and Manipulation (Ministry of Education), Renmin University of China, Beijing 100872, China}
\author {Xu-Jie Wang}
\email{wangxj@baqis.ac.cn}

\author {Li Liu}
\affiliation{Beijing Academy of Quantum Information Sciences, Beijing 100193, China}

\author {Hanqing~Liu}

\author {Haiqiao~Ni}

\author {Zhichuan Niu}
\affiliation{\textcolor{black}{State} Key Laboratory of Optoelectronic Materials and Devices, Institute of Semiconductors, Chinese Academy of Sciences, Beijing 100083, China}
\affiliation{Center of Materials Science and Optoelectronics Engineering, University of Chinese Academy of Sciences, Beijing 100049, China}

\author {Carlos Antón-Solanas}
\affiliation{Departamento de Física de Materiales, Instituto Nicolás Cabrera, Universidad Autónoma de Madrid, 28049 Madrid, Spain}
\affiliation{Condensed Matter Physics Center (IFIMAC), Universidad Autónoma de Madrid, Madrid 28049, Spain}

\author {Bang Wu}

\author{Zhiliang~Yuan}
\email{yuanzl@baqis.ac.cn}
\affiliation{Beijing Academy of Quantum Information Sciences, Beijing 100193, China}

\date{\today}% It is always \today, today,

\begin{abstract}%Number of words:141<150

Entanglement is a central resource in quantum technologies, and the realization of photonic entanglement necessarily relies on interaction with matter. Resonance fluorescence (RF), originating from the coherent interaction between a driving field and a two-level system, plays a pivotal role in quantum optics. Here, we demonstrate a novel route to entanglement generation based on RF from a single quantum dot. Rather than relying on generation of multiphoton states, 
our approach directly exploits vacuum–one-photon number superposition states created under resonant excitation. 
By delocalizing this superposition via a beam splitter, we realize time-bin entanglement and observe a clear violation of the Clauser–Horn–Shimony–Holt Bell inequality using Franson-type interferometry.
Our scheme removes the need for multiphoton generation, simplifies the experimental requirements, and establishes a scalable pathway toward solid-state entangled photon sources.

%Keywords: Entanglement, Resonance fluorescence, Bell inequality, quantum superposition states, interference 

\end{abstract}
\maketitle

Photonic quantum states underpin quantum information technologies, with their nonclassical features enabling computation, communication, and precision measurement beyond classical limits~\cite{Horodecki2009RMP,Chang2014,Yu2023,Wang2025}. Among these, superpositions of photon-number (Fock) states constitute one of the most essential quantum resources, since it provides an unbounded Hilbert space \cite{Rivera-Dean2022PRA,Stammer2022PRL}. 
While multi-photon Fock states ($\ket{n}$, with $n{>}1$) are often viewed as essential resources to explore the qudit encoding, the first step is to harness the superposition of vacuum and single-photon states,  $\sqrt{p_0} \ket{0}+ \sqrt{p_1} \ket{1}$. %, \zy{where \zy{$\alpha^2 + \beta^2 = 1$}}. 
Such states not only provide the basis for optical qubits but, more importantly, can be extended into highly efficient entangled states simply by correlating them across multiple spatial or temporal modes~\cite{Peiris2017PRL,Barkemeyer2021PRA}. Demonstrating this entanglement scheme offers a simpler and more efficient pathway than relying on the preparation of complex multi-photon states.

\begin{figure*}[hbt]
\centering 
\includegraphics[width=0.9\textwidth]{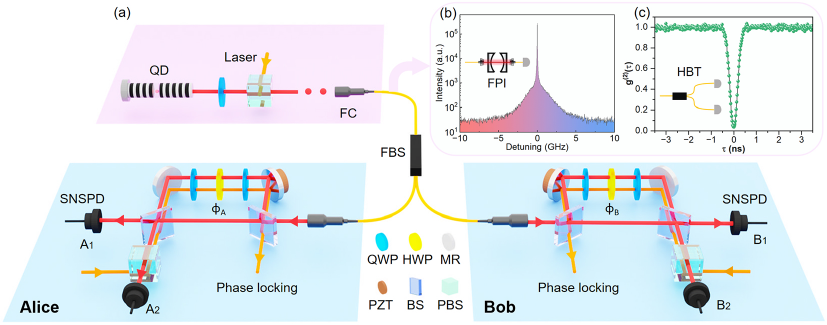}
\caption{ \textbf{Experimental setup.}  
\textbf{(a)}, The setup is composed by a quantum light source (light pink area) with a QD-micropillar device, resonantly driven by a CW laser, same resonant scheme as in Ref.~\cite{wu2023mollow}. The RF signal is split by a fiber beam splitter (FBS) and directed into two independent time-bin analyzers consisting of two AMZIs located at A and B sites(light blue areas). Coincidence counts are recorded by single-photon detectors (see detectors $A_1$, $A_2$ and $B_1$, $B_2$, respectively). Independent $\phi_A$ and $\phi_B$ phases of the two AMZIs are stabilized to desired values using a PID feedback loop based on the intensity of the phase locking laser. \textbf{(b)}, High-resolution RF spectra measured with a Fabry-P\`erot interferometer (FPI). \textbf{(c)}, Auto-correlation function $g^{(2)}(\tau)$ measured at an average photon number $\bar{n} = 0.01$ using a Hanbury Brown-Twiss (HBT) setup. FC, fiber collimator; QWP, quarter-wave plate; HWP, half-wave plate; MR, mirror; PZT, piezoelectric transducer; BS, beam splitter; PBS, polarizing beam splitter. $A_1$, $A_2$, $B_1$, and $B_2$ are the labels for the superconducting nanowire single photon detectors (SNSPDs).}
\label{fig1}
\end{figure*}  

In general, efficiently preparing and manipulating vacuum-one-photon states requires sophisticated quantum state engineering and conditional preparation~\cite{Qiao2021PRA,Das2022PRAppl,loredo_generation_2019}. Resonance fluorescence (RF) from a two-level quantum emitter offers a natural platform for preparing these superpositions. By resonantly exciting a quantum dot (QD), it has been shown the precise superposition control over $\ket{0}$ and $\ket{1}$ states, marking a significant breakthrough in quantum state preparation encoded in the photon-number basis~\cite{loredo_generation_2019,BuyWenniger2024,Karli2024,wang2025coherence}. Building on this scheme, the deterministic generation of time-entangled states via sequential pulsed driving, the teleportation of vacuum-one-photon states~\cite{Polacchi2024}, and the generation of two-photon states $\ket{2}$ via phase-controlled quantum interference~\cite{wang2025coherence,kim2025Arxiv} has been recently demonstrated. However, these efforts have not been systematically extended to distributed entangled-state between different spatial modes. In parallel, pioneering work has demonstrated energy–time entanglement between photon pairs harnessing the RF from natural and artificial atoms~\cite{liu2024violation,wang2025purcell}. Yet, these approaches rely on post-selecting two-photon components (e.g., $\ket{2}$ or sideband photon pairs), which severely limits efficiency.

In this work, we establish and experimentally demonstrate an alternative route: generating time-bin entanglement from vacuum–one-photon superposition states directly arising from the atom RF, without further preparing or filtering two-photon states. Using a Franson-type interferometer to implement projective measurements in the time-basis, we observe a clear violation of the Clauser–Horn–Shimony–Holt (CHSH) Bell inequality~\cite{Clauser1969PRL}, providing direct evidence of genuine entanglement shared between separate nodes.
%unambiguously confirming the entangled nature of the generated state between \st{Alice and Bob} \zy{separate} nodes. 
All experimental results are consistently explained and quantitatively reproduced within the framework of %our 
the pure-state model~\cite{wang2025coherence}. Our approach avoids reliance on multi-photon emission, and demonstrates the state-of-the-art efficiency of solid-state-based photon sources.

Figure~\ref{fig1} describes the experimental setup, following the standard Franson interferometer scheme~\cite{franson1989bell}. The quantum light source is composed of a self-assembled %semiconductor 
InAs QD coupled to a micropillar cavity, resonantly driven by a continuous wave (CW) laser. The excitation scheme is the same as in Ref.~\cite{wu2023mollow}; the laser and QD emission are not in cross-polarization configuration (see Fig.~\ref{fig1}(a)): the micropillar is engineered to be almost non-reflective at the cavity resonance. Hence, at low drive, the QD response dominates over the negligible cavity background (see Section II-1 of the Supplemental Material for further details). As a result of the RF excitation at low driving, the device emits a CW stream of single photons, see QD spectrum and anti-bunched photon statistics under a laser excitation of $\bar{n} = 0.01$ in Figs.~\ref{fig1}(b,c), respectively. The excitation fluxes ($\bar{n}$) were strictly calibrated via the incident optical power ($P_{in}$) upon the sample surface using the relation $\bar{n}  = P_{in}T_1/h\nu$, where $T_1$=67.2 ps is the exciton lifetime and $\nu=329.14$~THz is the pump laser frequency. For the sake of simplicity, and representing a good description of our single photon source emission at low driving, we describe the quantum light state as a pure superposition of temporally de-localized vacuum and single photons states $\sqrt{p_0}\ket{0}+\sqrt{p_1}e^{i \omega_0 t}\ket{1}$, where $p_0 + p_1 = 1$ and the one-photon mode takes the excitation laser phase~\cite{loredo_generation_2019,wang2025coherence}.  %the excitation laser phase \zy{($\omega_0 t$)} is taken as reference~\cite{loredo_generation_2019,wang2025coherence}.

In order to test the violation of the Bell inequality, a time-entangled state is generated. Initially, the vacuum-one-photon superposition stream is entangled between two spatial modes (Alice and Bob) after a fiber beamsplitter (FBS), see central part of Fig. \ref{fig1}. Vacuum-one-photon states from two different time bins, early ($e$) and late ($l$), form a time-bin entangled state resembling $1/\sqrt{2}(\ket{1^e_A,1^l_B}+\ket{1^l_A,1^e_B})$ (the factor $p_1/2$ is removed in this expression from the state post-selection). Then,  Alice and Bob implement local measurements in two independent asymmetric Mach-Zehnder interferometers (AMZIs)~\cite{takesue_implementation_2009}. The AMZIs temporal delay is $\tau=1.07$~ns (satisfying the condition $T_1\ll \tau \ll T_L$, where $T_L \approx 10$ $\mu$s is the laser coherence time). Independent projective measurements are performed in the Alice and Bob nodes by varying the phases $\phi_{A/B}$, and the outcomes are detected by the single-photon detectors $A_{1,2}/B_{1,2}$, respectively (see Fig.~\ref{fig1}).
The whole system is actively phase locked using the driving laser as the locking reference (orange beam in Fig. \ref{fig1}, see Section II-2 of the Supplemental Material for further details). 

\begin{figure*}[hbt]
\centering 
\includegraphics[width=1\textwidth]{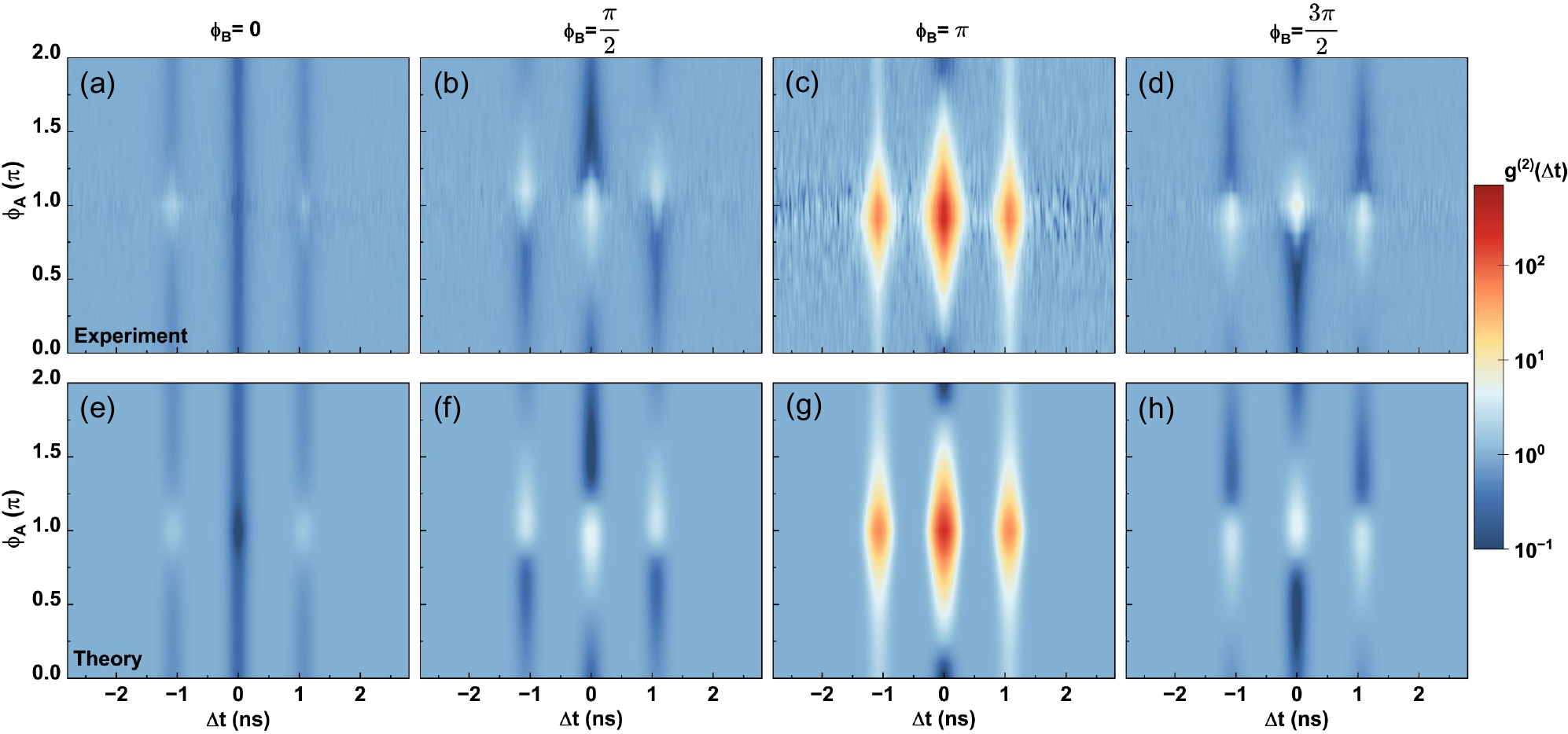}
\caption{\textbf{Mappings of the second-order correlation function $g^{(2)}(\Delta t)$}.  The laser driving power was set at $\bar{n} = 0.01$.
\textbf{(a-d)}, Normalized second-order correlation measurements of resonance fluorescence after the Franson interferometer between detectors $A_1$ and $B_1$, with $\phi_B$ fixed at 0, $\pi/2$, $\pi$, and $3\pi/2$, respectively, \zy{each $g^{(2)}(\Delta t)$ trace acquired with an integration time of four minutes}.  \textbf{(e-h)}, Theoretical simulations corresponding to panels \textbf{(a-d)}, respectively. The color intensity in the figure indicates the coincidence count rate at different phases, with all counts normalized to the maximum value. The color scale thus provides a direct visual representation of the variation in interference visibility.
}
\label{fig2}
\end{figure*}  

\begin{figure*}[hbt]
\centering 
\includegraphics[width=0.85\textwidth]{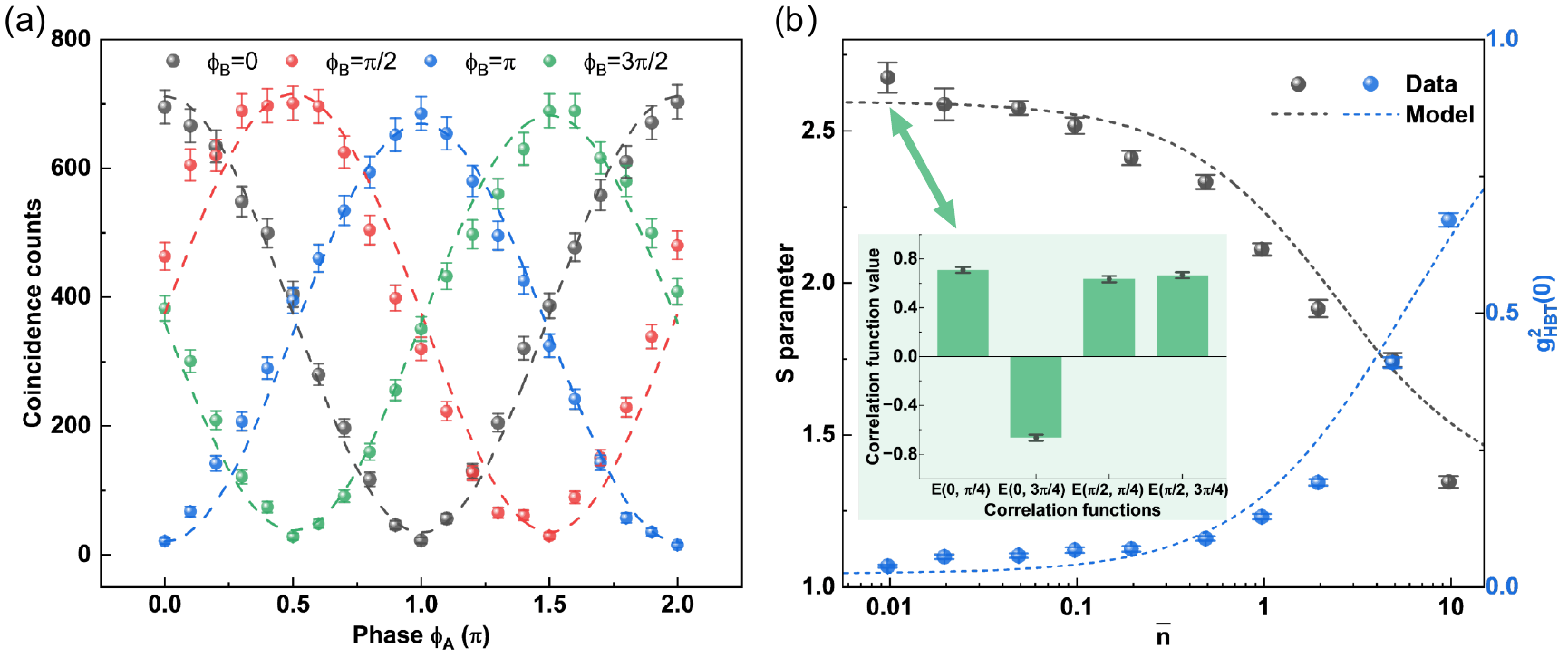}
\caption{ \textbf{Franson interference fringes %visibility of resonance fluorescence 
and observation of a CHSH inequality violation.}  
\textbf{(a)}, Correlation function $N$ for simultaneous coincidences in $A_1$ and $B_1$ as a function of $\phi_A$ for different values of $\phi_B$ = 0, $\pi$/2, $\pi$, 3$\pi$/2. Each point indicates the coincidence counts accumulated over a four-minute interval for a given phase setting. Dashed lines are fits using $N(\phi_A,\phi_B)=N_1\left[1+\cos(\phi_A-\phi_B)\right]+N_2$, with $N_1$, $N_2$ being fitting parameters. The average visibility of four correlation lines is 92.8 $\pm$ 2.6$\%$, which indicates the presence of time-entanglement. \textbf{(b)}, Pump power dependence of the S parameter and $g^{(2)}(0)$, with the dashed line denoting the theoretical prediction. The inset shows the measured correlation functions at an average input photon number $\bar{n} = 0.01$.}
\label{fig3}
\end{figure*}

%\zy{Z Zeng/J Wang: Write description of Fig.~2. D/L: 17/09/2025.}
Taking advantage of the flexible phase-control capability provided to %by 
the two AMZIs in Fig.~\ref{fig1}(a), we conduct for the first time a detailed phase-dependent Franson interference experiment, looking at the correlations at different time bins and local phases $\phi_{A/B}$. In the measurements, the phase of one AMZI, $\phi_B$, is fixed at values 0, $\pi/2$, $\pi$, and $3\pi/2$, while the phase of the other AMZI, $\phi_A$, is systematically scanned from 0 to $2\pi$ with a step size of $0.1\pi$. For each configuration of $\phi_A$ and $\phi_B$, we record the coincidence counts between single-photon detectors $A_1$ and $B_1$. The two-dimensional mappings of two-photon correlation are shown in Figs.~\ref{fig2}(a–d) for four different values of $\phi_B$. Here, the laser driving is $\bar{n}=0.01$ (low drive regime), and in each mapping, the data are independently normalized to the maximum coincidence counts within the corresponding dataset. From the experimental results at zero time delay, it can be observed that the two-photon coincidence counts exhibit a clear periodic dependence on $\phi_A$, with the interference fringes shifting as $\phi_B$ varies (see Fig.~\ref{fig3}(a) for details).

In the theoretical simulations, we employ the pure-state model~\cite{wang2025coherence} to accurately predict such correlations:

\begin{equation}    |\psi\rangle_t=\sqrt{p_0}|0\rangle_t|g\rangle_t+ \sqrt{p_1} e^{-2\pi i\nu t} \frac{|0\rangle_t|e\rangle_t+|1\rangle_t|g\rangle_t}{\sqrt{2}},
\label{PureState}
\end{equation}
where $\nu$ is the frequency of the CW laser, $\sqrt{p_0}$ and $\sqrt{p_1}$ are the magnitudes of the quantum probability amplitudes of vacuum and one-photon states, and $p_0+p_1=1$. Based on this state, we derive the expected coincidence count rates at different delays of $\Delta t$. For example, the central, simultaneous coincidence rate $\mathcal{C}(0)$ at 0-delay, in absence of optical losses, takes the form,
\begin{equation}
  \mathcal{C}(0)=\frac{p_1^2}{128}\left[ 1+\cos(\phi_A-\phi_B) \right].
\label{CountCenter}
\end{equation}

\noindent %\st{See Supplemental Material (Section II) for the expression of the coincidence rates at the delay of the AMZIs ($\Delta t=\pm\tau$) and the uncorrelated coincidence counts for other temporal delays.} 
%The detailed derivations of the expressions \zy{for the coincidence count rates} can be found in Section I of the Supplemental Material. 
The expressions for the coincidence count rates, including those at the AMZI delays ($\Delta t = \pm\tau$) and the uncorrelated coincidences at other temporal delays, together with their detailed derivations, can be found in Section~I-1 of the Supplemental Material. With the aid of the $g^{(2)}(t)$ function, the coincidence rates at the characteristic time delays can be sequentially combined (see Section I-2 of the Supplemental Material for details) to yield the theoretical predictions under the phase conditions corresponding to Figs.~\ref{fig2}(a–d), which are shown in Figs. 2(e–h). It is evident that this theoretical model faithfully reproduces the experimental results.

From the Bell inequality experiment of Fig.~\ref{fig1}, we construct and measure the $S$ function that allows to verify the time-bin entanglement of the vacuum-one-photon superposition state distributed to Alice and Bob nodes. This $S$ function is defined as $S=\left|  E(\phi_A,\phi_B)- E(\phi_A,\phi'_B)+ E(\phi'_A,\phi_B)+ E(\phi'_A,\phi'_B) \right|$, where the set of %\st{angles} 
phase delays of Alice and Bob are $\{\phi_A,\phi'_A\}{=}\{0,\pi/2\}$ and $\{\phi_B,\phi'_B\}{=}\{\pi/4,3\pi/4\}$. 
$E(\phi_A,\phi_B)$ denotes the correlation function, %\st{under the corresponding phase combination}, 
which can be 
computed %obtained 
using the relation~\cite{Clauser1969PRL,liu2024violation,wang2025purcell},
\begin{equation}
\begin{aligned}
&E(\phi_A,\phi_B)=\\
&\frac{N(\phi_A,\phi_B)+N(\phi_A^\perp,\phi_B^\perp)-N(\phi_A^\perp,\phi_B)-N(\phi_A,\phi_B^\perp)}{N(\phi_A,\phi_B)+N(\phi_A^\perp,\phi_B^\perp)+N(\phi_A^\perp,\phi_B)+N(\phi_A,\phi_B^\perp)},
\end{aligned}
\end{equation}
where $N(\phi_A, \phi_B)$ denotes the measured coincidence count for $\Delta t = 0$ under the phase setting ($\phi_A, \phi_B$), and $\phi^\perp=\phi+\pi$. Experimentally, the output ports $A_1/B_1$ and $A_2/B_2$ exhibit a relative phase shift of $\pi$, such that  coincidences measured simultaneously among the four output combinations ($A_1,B_1$), ($A_1,B_2$), ($A_2,B_1$), and ($A_2,B_2$) enable an efficient determination of the correlation function $E(\phi_A,\phi_B)$.

Figure \ref{fig3}(a) shows the coincidence counts $N(\phi_A,\phi_B)$ at $\Delta t = 0$ between detectors $A_1$ and $B_1$ as a function of $\phi_A$ for 4 different values of $\phi_B$. According to Eq.~\ref{CountCenter}, $N(\phi_A,\phi_B)$ is proportional to $1+\cos(\phi_A-\phi_B)$. 
The corresponding 16 correlations of $N(\phi_A,\phi_B)$ enable the measurement of the previously defined $E(\phi_A,\phi_B)$ functions, from which the $S$ parameter can be computed. %reconstructed.
Figure \ref{fig3}(b) shows the $S$ parameter (black dots) and the second-order correlation function ($g^{(2)}_{\textrm{HBT}}(0)$, blue dots) as a function of $\bar{n}$. An increasing laser driving power (proportional to $\bar{n}$) enhances the presence of unrejected laser light in the collection, and so reduces the $S$ parameter. For low driving ($\bar{n}{\sim}0.01$, and $g^{(2)}_{\textrm{HBT}}(0){\sim}0.037(3)$), we observe $S=2.675(50)>2$, which exceeds the classical bound 2 by more than 13 standard deviations.
The inset in Fig.~\ref{fig3}(b) shows the four $E(\phi_A,\phi_B)$ correlation functions for $\bar{n}{\sim}0.01$ driving. %\st{, from which the corresponding $S$ function is computed.} 

This significant Bell inequality violation provides strong evidence of time-bin entanglement between Alice and Bob nodes, confirming the nonlocal nature of the generated quantum state. As the pump power gradually increases from $\bar{n}$ = 0.01 to 9.81, the $g^{(2)}_{\textrm{HBT}}(0)$ also increases up to 0.67(1), \xj{mainly due to the influence of laser background noise~\cite{wu2023mollow},} where the $S$ parameter lies below the violation limit $S\sim1.344(19)$.  The crossover of the Bell inequality violation ($S=2$) occurs for a $g^{(2)}_{\textrm{HBT}}(0)$ value of approximately 0.2, where we estimate that the unwanted laser component of the collected light constitutes $43\%$ of the total average photon number~\cite{stevens2014third,2016A}. 

We note that while there have been recent reports on generating similar entangled states using single-atom RF~\cite{hu2025transforming}, our source exhibits a brightness exceeding such previous implementations by more than two orders of magnitude, even under the weakest excitation conditions in Fig. \ref{fig3}(b), highlighting the efficiency of QD-micropillar sources. It is also worth noting that the entanglement demonstrated here differs from other schemes that also generate entangled photons via RF~\cite{liu2024violation,wang2025purcell}. In those experiments, the entanglement is extremely sensitive to excitation power, only present at sufficiently weak excitation powers. In contrast, the entanglement quality in our case, where no cross-polarisation excitation scheme is followed, is only tied to the single-photon purity and the micropillar cavity design. Cross-polarized RF excitation would further increase the region where $S>2$ for higher $\bar{n}$ values.

\begin{figure*}[hbt]
\centering 
\includegraphics[width=0.8\textwidth]{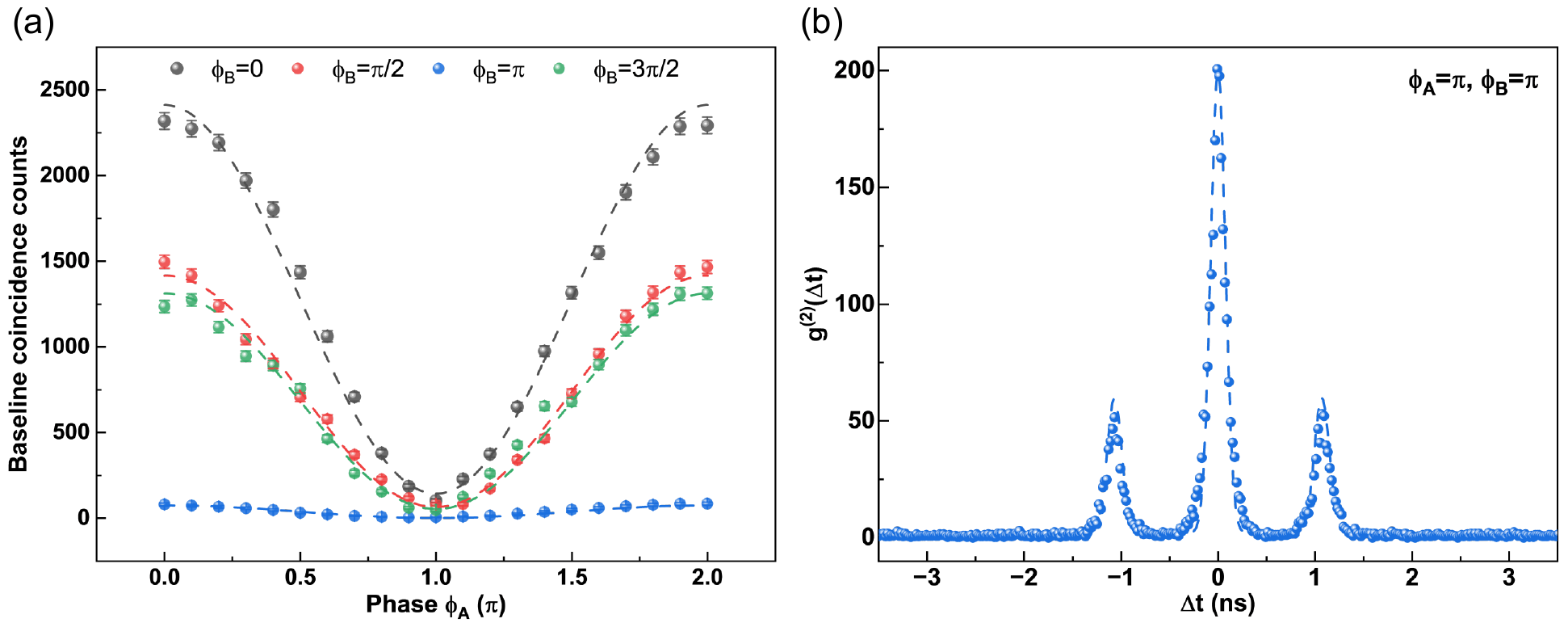}
\caption{\textbf{Effect of the first-order interference.} \textbf{(a)}, Baseline coincidence counts as a function of $\phi_A$ for different $\phi_B$ values, measured at the laser driving power of $\bar{n} = 0.01$. 
\textbf{(b)}, Experimental second-order correlation measured under the phase configuration $(\phi_A,\phi_B) = (\pi, \pi)$. 
In both panels, the dots represent the experimental data, while the dashed lines show simulations obtained using the same calculation method as in Fig.~\ref{fig2}(b).}
\label{fig4}
\end{figure*}

Finally, we \zy{revisit the raw data of Figs.~\ref{fig2}(a–d) to examine how the coherence of vacuum–one-photon number superposition states manifests in the Franson interferometry.
Figure~\ref{fig4}(a) derives from these data and shows the baseline coincidence count as a function of $\phi_A$ for four representative values of $\phi_B$.}
%Figure~\ref{fig4}(a) shows the baseline coincidence count as a function of $\phi_A$ for four different values of $\phi_B$, measured with an integration time of four minutes. 
The data are well described by $\mathcal{C}_0 \propto (1 + V_A \cos\phi_A)(1 + V_B \cos\phi_B)$, where $V_A$ and $V_B$ denote the interference visibilities. 
The baseline coincidence reaches a minimum when both AMZIs are set to destructive interference ($\phi_{A,B} = \pi$). 
In contrast, the zero-delay coincidence arises from two-photon interference and depends solely on the phase difference $\phi_A - \phi_B$, being insensitive to the individual AMZI phases. 
This difference in phase dependence enables continuous tuning of the correlation between detectors $A_1$ and $B_1$ from anti-bunching to super-bunching. 
As an illustration, Fig.~\ref{fig4}(b) shows the strongest bunching observed at $(\phi_A,\phi_B) = (\pi,\pi)$. 
We note that such behavior is not observable in non-resonantly excited single-photon sources~\cite{bennett2008experimental}.

In conclusion, we have experimentally demonstrated that vacuum–one-photon number superposition states generated via resonant excitation of a QD can be harnessed to fundamentally test the CHSH Bell inequality, evidencing the generation of time-entanglement in a Franson-like interferometer. Compared with other RF-based approaches to generate entanglement, our scheme is simple to implement, requiring neither complex excitation nor signal post-processing, while offering low power consumption, high brightness, and excellent transmission stability for communication protocols. These advantages point to a promising route for transforming quantum entanglement from proof-of-principle demonstrations to practical, integrated implementations. Furthermore, our investigation of the connection between the spectral components of RF and the observed antibunching, together with the faithful reproduction of the experimental results by our simple pure-state model, provides deeper insight into the nature of atomic RF. 

Going beyond previous demonstrations of distributed entanglement among three time bins and two spatial modes~\cite{Thew2004}, or the converse configuration of three spatial modes and two time bins~\cite{Agne2017}, the high brightness of our source opens the way to extending time-bin entanglement across multiple spatial modes, thereby scaling nonlocal tests in a straightforward manner. For example, by replacing the fiber beam splitter with a tritter~\cite{Spagnolo2013} or a 1-to-3 fiber coupler, and injecting three single-photon states in the temporal modes \textit{early} (\(e\)), \textit{middle} (\(m\)), and \textit{late} (\(l\)), one can post-select the tripartite entangled qutrit state $\frac{1}{\sqrt{6}}\sum_{\pi \in S_3}\big|t_{\pi_1}\big\rangle_A\,\big|t_{\pi_2}\big\rangle_B\,\big|t_{\pi_3}\big\rangle_C$, where \(S_3\) is the symmetric group of degree three, containing all \(3! = 6\) permutations of \((e,m,l)\) timebins in the three \(A,B,C\) spatial modes. This state comprises three photons occupying distinct temporal and spatial modes, and its post-selection probability scales as $n!/n^n$ ($\sim 0.22$ for $n=3$), suggesting that the distribution of high-dimensional, genuinely multipartite entanglement among a larger number of modes is within reach for QD-micropillar devices in forthcoming implementations \cite{Martin2017,Caspar2022}.

\vspace{0.2 cm}
\textit{Acknowledgments}---This work was supported by the National Natural Science Foundation of China under grants (12494600, 12494604, 12204049, 12274223 and 12504410) and the Beijing Postdoctoral Research Foundation. C. A-S. acknowledges the support from the projects from the Ministerio de Ciencia e Innovaci\'on PID2023-148061NB-I00 and PCI2024-153425, the project ULTRABRIGHT from the Fundaci\'on Ram\'on Areces, the Comunidad de Madrid fund “Atracci\'on de Talento, Mod. 1”, Ref. 2020-T1/IND-19785, and the “María de Maeztu” Program for Units of Excellence in R\&D (CEX2023-001316-M).

\bibliography{myref}

~

~

\end{document}